\documentclass[apj,twocolumn]{emulateapj}

\slugcomment{accepted in ApJ}

\shortauthors{Parmentier et al.}
\shorttitle{Initial cluster mass function}

\begin{document}


\title{The shape of the initial cluster mass function:  \\
what it tells us about the local star formation efficiency}


\author{G.~Parmentier\altaffilmark{1,2,4}, S.~P.~Goodwin\altaffilmark{3}, P.~Kroupa\altaffilmark{1} and H.~Baumgardt\altaffilmark{1}}





\altaffiltext{1}{Argelander-Institut f\"ur Astronomie, University of Bonn,
Auf dem H\"ugel 71, D-53121 Bonn, Germany}
\altaffiltext{2}{Institute of Astrophysics \& Geophysics, University of Li\`ege, 
All\'ee du 6 Ao\^ut 17, B-4000 Li\`ege, Belgium}
\altaffiltext{3}{Department of Physics and Astronomy, University of Sheffield, 
Hicks Building, Hounsfield Road, Sheffield S3 7RH, United Kingdom}
\altaffiltext{4}{Alexander von Humboldt Fellow and Research Fellow of Belgian
Science Policy}


\begin{abstract}
We explore how the expulsion of gas from star-cluster forming cloud-cores due 
to supernova explosions affects the shape of the initial cluster mass 
function, that is, the mass function of star clusters when effects of 
gas expulsion are over. We demonstrate that if the radii of 
cluster-forming gas cores are roughly constant over the core 
mass range, as supported by observations, then more massive cores undergo 
slower gas expulsion. Therefore, for a given star formation
efficiency, more massive cores retain a larger fraction of stars after gas 
expulsion.  The initial cluster mass function may thus differ from the 
core mass function substantially, with the final shape depending on
the star formation efficiency.  A mass-independent star
formation efficiency of about 20\,per cent turns a power-law core 
mass function into a bell-shaped initial cluster mass function, 
while mass-independent efficiencies of order 40\,per cent preserve 
the shape of the core mass function.
\end{abstract}


\keywords{galaxies: star clusters --- stars: formation --- stellar dynamics}

\section{Introduction}

One of the greatest discoveries made by the {\sl Hubble Space
Telescope} is that  the formation of star clusters with mass and
compactness comparable to those  of old globular clusters is still
ongoing in violent star forming environments,  such as starbursts and
galaxy mergers.  The young cluster mass function (CMF: the number
of objects per logarithmic cluster  mass interval, ${\rm d}N/{\rm
dlog}m$) is often reported to be a power-law of  spectral index
$\alpha =-2$ (i.e. equivalent to ${\rm d}N \propto m^{-2} {\rm d}m$)
down to the detection limit~\citep{fal01, bik03, hun03}.  Similar
results are obtained  for the cluster luminosity
function~\citep[e.g.,][]{mil97,whi99}.  However, old globular  cluster
systems are found to have a CMF which is Gaussian-like with a peak
mass of  $\sim 10^5\,{\rm M}_{\odot}$ \citep{kav97}.  These very
different CMFs between young  and old systems present the greatest
dissimilarity between the two populations, and  perhaps the greatest
barrier to identifying young massive clusters as analogues of  young
globular clusters.

Yet, it is worth keeping in mind that the luminosity/mass distribution of
some young star cluster systems does {\sl not} obey a power-law.  
Based on high resolution {\sl Very Large Array} observations,
\citet{joh08} rules out a power-law cluster luminosity function for
the young star clusters formed in the 
starburst galaxy Henize 2-10.  In NGC~5253, an irregular 
dwarf galaxy in the Centaurus Group, \citet{cre05} detect a turnover at 
$5 \times 10^4\,{\rm M}_{\odot}$ in the mass function of the young 
massive star clusters in the central region.
Last but not least, the case of the Antennae merger
of galaxies NGC~4038/39 remains heavily debated.  While \citet{whi99} infer 
from their {\sl HST/WFPC2} imaging data a power-law cluster luminosity function 
with a spectral index $\alpha \simeq -2.1$, \citet{and07} report the first 
statistically robust detection of a turnover at $M_V \simeq -8.5$\,mag, 
corresponding to a mass of $\simeq 16 \times 10^3\,{\rm M}_{\odot}$.
The origin of that discrepancy likely resides in differences in the data 
reduction and the statistical analysis aimed at rejecting bright-star 
contamination and disentangling completeness effects from intrinsic cluster 
luminosity function substructures.

These findings raise the highly interesting prospect that the shape of the 
CMF at young ages does vary from one galaxy to another.  If it were conclusively 
proven that the shape of the initial cluster mass function (ICMF) is {\it not} 
universal among galaxies but, instead,
depends on environmental conditions and/or on the star formation process, 
consequences would be far reaching since this would imply that the 
young CMF may help us decipher physical conditions prevailing in external
galaxies.

In this contribution, we report the first results of a project investigating 
the evolution of the mass function of star forming cores into that of
bound gas-free star clusters which have survived expulsion of their
residual star forming gas.  A similar study was carried out by \citet{par07}
in which they showed that protoglobular cloud mass functions characterized by
a mass-scale of $\simeq 10^6\,{\rm M}_{\odot}$ evolve into the observed globular
cluster mass function.  They propose that the origin of the universal 
Gaussian mass function 
of old globular clusters thus likely resides in a protoglobular cloud 
mass-scale common among galaxies and imposed in the protogalactic era. 
Present-day star forming cores, however, do not show any characteristic mass-scale
as their mass function is a power-law down to the detection limit.  
Previously, Kroupa \& Boily (2002) showed that a power-law embedded-cluster 
mass function can evolve into a mass function with a turnover near 
$10^5\,{\rm M}_{\odot}$ if the physics of gas removal is taken into account.
Taking advantage of the grid of $N$-body models recently compiled by 
\citet{bau07}, we explore in greater detail how a featureless core mass 
function evolves into the ICMF as a result of gas removal, which we model as 
the supernova-driven expansion of a supershell of gas.  We also explore 
how the ICMF responds to model parameter variations so as to understand 
what may make the mass function of young star clusters vary from one galaxy
to another.  In a companion paper \citep{bau08}, we explore this 
issue from a different angle, that is, by defining an energy criterion, and 
we apply that alternative model to the mass function of the Milky Way old
globular clusters.

\section{Notion of Initial Cluster Mass Function}
\label{sec:icmf}

Before proceeding any further, it is worth defining the notion of the
ICMF properly.  Following the onset of massive star activity at an age
of 0.5 to 5\,Myr, a gas-embedded cluster expels its residual star
forming gas.  As a result, the potential in which the stars reside
changes rapidly causing the cluster to violently relax in an attempt
to reach a new equilibrium. This violent relaxation takes 10 -- 50~Myr
during which time the cluster loses stars ("infant weight-loss") and
may be completely destroyed ("infant-mortality") (this process has
been studied in detail by \citet{hil80}; \citet{mat83}; \citet{elm83};
\citet{lad84}; \citet{elm85}; \citet{pin87}; \citet{ver89};
\citet{goo97a}; \citet{goo97b}; \citet{kah01}; \citet{gey01}; \citet{boi03a};
\citet{boi03b}; \citet{goo06}; \citet{bau07}).  After $\sim 50$~Myr
the evolution of clusters is driven by internal two-body relaxation
processes and tidal interactions with the host galaxy
\citep[see][]{spi87}. Following \citet{kro02} we define the ICMF 
as the mass function of clusters {\em at the end of the violent relaxation phase}, 
that is, when their age is $\simeq 50$\,Myr.

The effect of gas expulsion depends strongly on the virial ratio of
the cluster {\em immediately before} the gas is removed.  If the stars
and gas are in virial equilibrium before gas expulsion the virial
ratio $Q_\star$ of the stars depends entirely on the star formation
efficiency (SFE, or $\varepsilon$) such that $Q_\star = 1/(2\varepsilon)$
(where we define virial equilibrium to be $Q=1/2$) \citep{goo08}.  
Formally, $\varepsilon$ is the {\it effective} SFE (eSFE: see \citet{ver90};
\citet{goo06}), that is, a measure of how far from virial equilibrium
the cluster is at the onset of gas expulsion, rather than the actual
efficiency with which the gas has formed stars.  We note that even
quite small deviations from virial equilibrium can equate to large
differences between the eSFE and the SFE.  However, in this paper we
will make the assumption that the stars have had sufficient time to
come into virial equilibrium with the gas potential (ie. gas expulsion
occurs after a few crossing times), and so the eSFE very closely
matches the SFE.  We discuss this assumption further in the
conclusions.

In this case, the radius of the embedded cluster is the same
as that of the core and the three-dimensional velocity dispersion 
of the stars in the embedded cluster is dictated by the depth of
the core gravitational potential, namely: 
$\sigma = (G m_{\rm c}/r_c)^{1/2}$, where $m_c$ and $r_c$ are the 
mass and radius of the core, respectively, and G is the gravitational
constant.

A star-cluster forming `core' of mass $m_{\rm c}$ (ie. the region of a
molecular cloud that forms a cluster) with a SFE of $\varepsilon$
will form a cluster of mass $\varepsilon \times m_{\rm c}$.  However,
gas expulsion will unbind a fraction $(1 - F_{\rm bound})$ of the
stars in that cluster such that the mass of the cluster at $\sim
50$~Myr will be
\begin{equation} 
m_{\rm init} = F_{\rm bound} \times \varepsilon \times m_{\rm c}\;.
\label{eq:minit}
\end{equation}
(Note that when $F_{\rm bound} \sim$~zero the cluster is destroyed.)

It is important to keep in mind that, for clusters whose age does not
exceed,  say, 10-20\,Myr, the mass fraction of stars that have escaped
after gas removal  is still small, as most stars do not have a high
enough velocity to have become spatially dissociated from the cluster
(see fig.~1 in \citet{goo06} and fig.~1 in \citet{kah01}).  The
instantaneous mass of these young clusters thus follows:  $m_{\rm cl}
\lesssim \varepsilon \times m_{\rm c}$ and, in the absence of  any
explicit dependence of $\varepsilon$ on $m_{\rm c}$, the CMF at such a
young age mirrors the core mass function.   Any meaningful comparison
of our model ICMFs therefore has to be limited to populations of
clusters which have either re-virialised or been destroyed.

In our Monte-Carlo simulations, we assume that the SFE of a core is
independent of its mass.  The core mass function is assumed to follow
a  power-law of spectral index $\alpha =-2$ and we describe the
probability  distribution function of the SFE as a Gaussian of mean
$\bar\varepsilon$  and standard deviation $\sigma _\varepsilon$, which
we denote  $P(\varepsilon)=G(\bar\varepsilon, \sigma _\varepsilon)$.
  
Clearly, if the bound fraction of stars after gas expulsion, $F_{\rm
  bound}$, depends upon the core mass, then the shape of the ICMF will
  differ from the (power-law) core mass function, an effect whose
  study was  pioneered by \citet{kro02}.  We now take this point
  further by combining the detailed  $N$-body model grid of
  \citet{bau07}, which provides the bound fraction  $F_{\rm bound}$ as
  a function of the SFE $\varepsilon$ and of  the ratio between the
  gas removal time-scale $\tau _{\rm GR}$ and the protocluster
  crossing-time $\tau _{\rm cross}$.

\section{The gas expulsion time-scale}
\label{sec:grts}

We model the gas expulsion from an (embedded) cluster as an outwardly
expanding supershell whose radius $r_s$ varies with time $t$ as
\citep{cas75}:
\begin{equation}
r_s(t)=\left( \frac{125}{154 \pi} \frac{\dot{E}_0}{\rho _g}
\right)^{1/5} t^{3/5}\;.
\label{eq:rs}
\end{equation}
Here $\dot{E}_0$ is the energy input rate from Type II supernovae and
$\rho _g$ is the mass density of the residual star forming gas in the
core.  We assume the mass density profiles of the core and of the
newly formed stars to be alike, that is, $\rho _g =
(1-\varepsilon)\rho _c$.

The energy input rate from massive stars expelling the gas is
\begin{equation}
\dot{E}_0 = \frac{N_{\rm GR} \times E_{\rm SN}}{\tau _{GR}}\,,
\label{eq:dotE_GR}
\end{equation}
where $N_{\rm GR}$ is the number of supernovae expelling the gas, 
$E_{\rm SN}$ is the energy of one single supernova, and $\tau _{GR}$ 
is the gas removal time-scale.  At this stage, however, we know neither 
$N_{\rm GR}$ nor $\tau _{GR}$.  Since the energy input rate from 
massive stars is approximately constant with time over the whole supernova 
phase, we thus rewrite:

\begin{equation}
\dot{E}_0 = \frac{N_{\rm SN} \times E_{\rm SN}}{\Delta t_{\rm SN}}\;.
\label{eq:dotE}
\end{equation}
$\Delta t_{\rm SN}$ is the supernova phase duration (i.e. the time from 
the first to the last supernova) and $N_{\rm SN}$ is the total 
number of supernovae formed in the core.  We caution that a fraction 
only of these contribute to gas expulsion, that is, 
$N_{\rm SN}>N_{\rm GR}$ and $\Delta t_{\rm SN}>\tau _{\rm GR}$.

The gas removal time-scale corresponds to the instant when the
supershell radius is equal to the core radius $r_c$, namely,
$r_s(t=\tau _{\rm GR})=r_c$.  The combination of that constraint on
$\tau _{\rm GR}$ with eqs.~\ref{eq:rs} and \ref{eq:dotE} gives:
\begin{equation}
\tau _{GR} = r_c^{2/3} \left( \frac{154}{125} \frac{3}{4} \,
\frac{\Delta t_{\rm SN}}{N_{SN} E_{\rm SN}} \,(1-\varepsilon ) m_c
\right)^{1/3}\;.
\label{eq:tauGR}
\end{equation}

While gas-embedded cluster masses vary by many orders of magnitude,
their radii are remarkably constant, always of order 1\,pc (see table
1 in \citet{kro05}).  As  $N_{SN} \propto m_c$, it follows from
eq.~\ref{eq:tauGR} that, for a given  $\varepsilon$, the gas removal
time-scale $\tau _{GR}$ is independent of the core mass $m_c$.
However, the impact of gas expulsion upon a cluster is not governed by
the absolute value of $\tau _{\rm GR}$, but  by its ratio with the
core crossing time $\tau _{\rm cross}$, ie.  $F_{\rm bound}$ is a
function of $\tau _{\rm GR} / \tau _{\rm cross}$.  The dynamical
crossing-time of the star forming core is given by  $\tau _{\rm cross}
= (15 / \pi G \rho _c )^{0.5}$ which, combined  with
eq.~\ref{eq:tauGR}, gives:

\begin{equation}
\frac{\tau _{\rm GR}}{\tau _{\rm cross}}  = 1.9 \times 10^{-4}  \left(
\frac{\Delta t_6}{E_{51}} \, \frac{1-\varepsilon}{\varepsilon}
\right)^{1/3}  \left( \frac{m_c}{1\,{\rm M}_\odot}\right)^{1/2}  \,
\left( \frac{r_c}{1\,{\rm pc}}\right)^{-5/6}\;.
\label{eq:ratio_tau}
\end{equation}

In this equation, the number of supernovae, $N_{SN}$, from
eq.~\ref{eq:tauGR} is expressed as a function of the embedded cluster
mass $m_{ecl} = \varepsilon \times m_c$,  assuming a Kroupa stellar
initial mass function \citep{kroupa01}.  We also assume that the upper
stellar mass limit correlates with $m_{ecl}$,  as found by
\citet{wei04}, although the hypothesis of a fixed value for the mass
of the most massive supernova progenitor (say, 60\,${\rm M}_{\odot}$)
hardly affects the results presented in the next section.  $E_{51}$
and $\Delta t_{6}$ are the energy released per supernova in units  of
$10^{51}$\,ergs and the duration of the supernova phase, in Myr,
respectively.  In what follows, we adopt $E_{51}=1$ and $\Delta
t_{6}=30$.

Eq.~\ref{eq:ratio_tau} shows that, in the absence of a mass-radius
relation, the gas removal time-scale, expressed in units of the core
crossing time, explicitly depends on the core mass.  For a given
$\varepsilon$, the deeper potential well of more massive cores slows
down gas expulsion when compared to low-mass cores.  Massive clusters
are therefore better able to adjust to the new gas-depleted
potential, and have a larger bound fraction $F_{\rm bound}$.
Eq.~\ref{eq:ratio_tau} is shown graphically in Fig.~\ref{fig:fbound}
as a match between bottom and top $x$-axes for $\varepsilon =0.33$ 
and $r_c=1$\,pc.
In the next section, we explore how the different ICMFs can be
produced from the same (power-law) core mass function.
 
\section{From the core mass function to the initial cluster mass function}
\label{sec:CMF}

The core mass function is modelled as a power-law of spectral index
$\alpha =-2$ (i.e. ${\rm d}N \propto m_c^{-2}{\rm d}m_c$) and the
distribution  function for the core (local) SFE is a Gaussian, as
defined in Section \ref{sec:icmf}.  Both distributions are sampled
independently,  that is, the SFE is assumed to be independent of the
core mass.  The core radius is set to be $r_c \simeq 1$\,pc and the
ratio  $\tau _{GR} / \tau _{cross}$ is given by eq.~\ref{eq:ratio_tau}.

The fraction $F_{\rm bound}$ of stars remaining bound to the cluster
after gas removal is inferred by linearly interpolating the grid of
\citet{bau07} (their table 1) which describes how $F_{\rm bound}$
varies with $\varepsilon$ and $\tau _{GR} / \tau _{cross}$.

\begin{figure}
\begin{center}
\epsscale{1.2}  \plotone{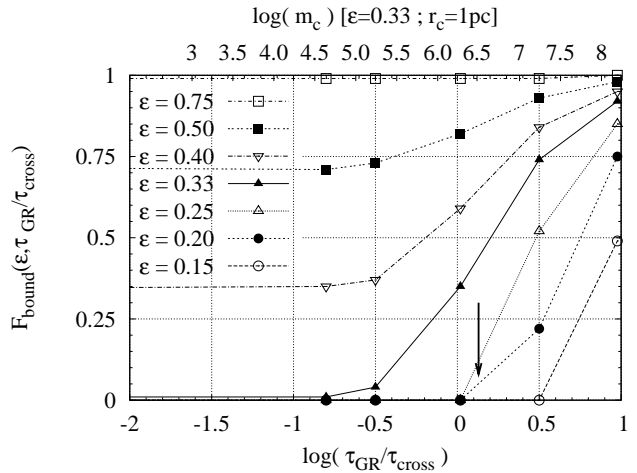}
\caption{Mass fraction $F_{\rm bound}$ of stars remaining bound to the
cluster at the end of the violent relaxation phase induced by gas
expulsion.   $F_{\rm bound}$ varies with the SFE $\varepsilon$ and
with the ratio between  the gas removal time-scale ${\tau _{GR}}$ and
the core crossing-time  ${\tau _{cross}}$.  The arrow indicates the
bound fraction $F_{\rm bound}$  and the core mass $m_c$ corresponding
to the ICMF turnover when the mean SFE is $\bar\varepsilon =0.20$ 
and the standard deviation of the SFE probability distribution is 
$\sigma_{\varepsilon}=0.00$ (also marked  by an arrow in the top panel 
of Fig.~\ref{fig:icmf})}
\label{fig:fbound}
\end{center} 
\end{figure}

According to our gas removal model, $\tau _{GR}$ is the time-scale
over which the cluster expels the entirety of its gas.  However,
\citet{bau07} model  gas removal as an exponential decrease with time
of the cluster gas content, and their  gas removal time-scale ($\tau
_M$ in their eq.~1) corresponds to the time when the residual gas has
a fraction $e^{-1}=0.37$ of its initial value.  Prior to using their
results, we multiply $\tau _M$  by a factor of 3 (i.e. $\tau
_{GR}=3\tau _M$), so that $\tau _{GR}$ now corresponds to a residual
gas mass fraction of $e^{-3}=0.05$, i.e., the cluster  is practically
devoid of gas.  The corresponding grid is shown in
Fig.~\ref{fig:fbound}.   In this paper, we focus on clusters evolving
in a low-density  environment/weak tidal field and we thus consider
the lowest half-mass  radius to tidal radius ratio in the grid of
\citet{bau07}, i.e. $r_h/r_t=0.01$.   In stronger tidal fields the
shape of the ICMF is likely to be more affected than what we find
below, since low-mass clusters have smaller tidal radii  and are,
therefore, more easily disrupted by gas expulsion.

\begin{figure}
\begin{center}
\epsscale{1.2} \plotone{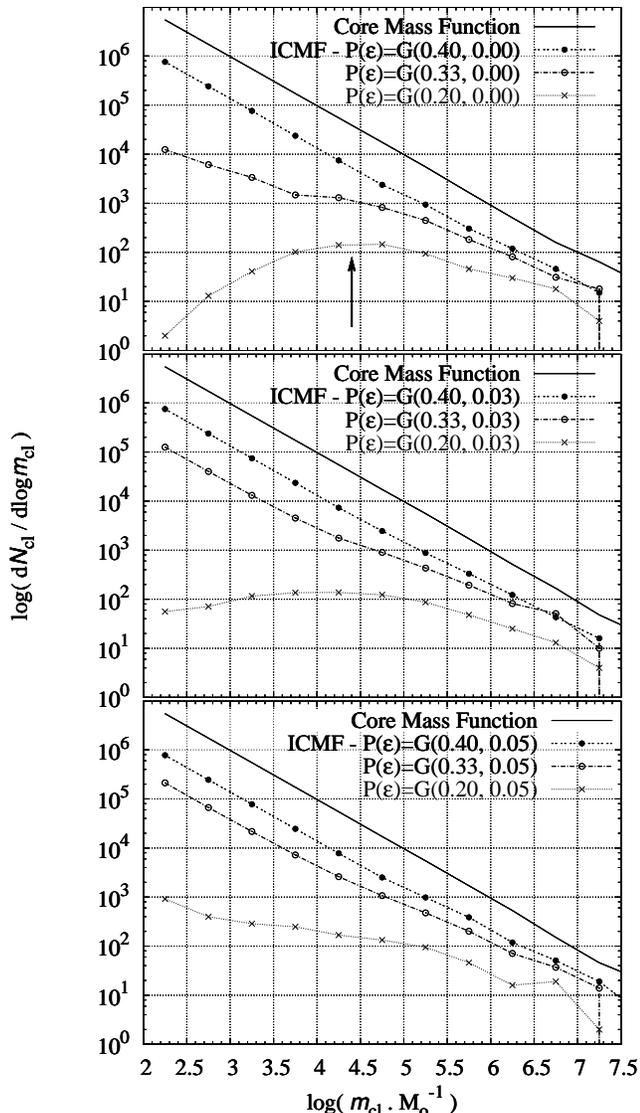}
\caption{How the form of the ICMF produced from a $-2$ power-law
  cluster mass function responds to variations in the SFE
Gaussian distribution G($\bar\varepsilon$, $\sigma _{\varepsilon}$),
where $\bar\varepsilon$  and $\sigma _{\varepsilon}$ are the mean SFE
and the standard deviation, respectively.   Solid lines depict the
original core mass function.}
\label{fig:icmf}
\end{center} 
\end{figure}

Our Monte-Carlo simulations, based on eq.~\ref{eq:minit} and shown in
Fig.~\ref{fig:icmf}, show the evolution of the star forming core mass
function into the ICMF.   We have considered 9 different cases,
combining three mean star formation efficiencies
($\bar\varepsilon=0.20, 0.33, 0.40$) with  three standard deviations
($\sigma _{\varepsilon}=0.00, 0.03, 0.05$).  Examination of the panels
of Fig.~\ref{fig:icmf} shows that depending on the assumed
distribution of SFEs $G(\bar\varepsilon, \sigma _\varepsilon)$,
markedly  different ICMFs are produced, ranging from a power-law shape
(e.g. when $\bar\varepsilon=0.40$, regardless of $\sigma
_{\varepsilon}$)  to a bell-shape (e.g. when $\bar\varepsilon=0.20$
and $\sigma _{\varepsilon} \lesssim 0.03$).
  
These different behaviours stem from how large the variations of the
bound fraction  $F_{\rm bound}$ with the core mass $m_c$ are.  In
order to illustrate this, let us focus on the top panel of
Fig.~\ref{fig:icmf} for which $\sigma _{\varepsilon}=0$.  A SFE as
high as $\varepsilon = 0.4$ guarantees that $F_{\rm bound}$  is only
a weak function of the core mass $m_c$ as $0.35  \lesssim F_{\rm
bound} \lesssim 0.95$ over the core mass range  $10^3\,{\rm M}_{\odot}
\lesssim m_c \lesssim 10^8\,{\rm M}_{\odot}$  (see
Fig.~\ref{fig:fbound}).  As a result, the shape of the ICMF mirrors
that of the core mass function and is close to a power-law of spectral
index $\alpha =-2$.

In the case of a SFE of $\bar\varepsilon =0.33$, the same core mass
range  corresponds to an increase of the bound fraction 
with the core mass of almost two
orders of magnitude  (i.e. $0.01 \lesssim F_{\rm bound} \lesssim
0.90$), making the ICMF significantly shallower than the core mass
function.  It is worth keeping in mind  that the detailed form of this
ICMF is uncertain, as the determination of the bound fraction $F_{\rm
bound}$ depends on the fine details of the $N$-body modelling of gas
expulsion when  $F_{\rm bound} \lesssim 0.1$.  As for $\varepsilon
=0.33$,  $F_{\rm bound} \simeq 0.01{\rm -}0.04$ when  $\log (\tau
_{GR} / \tau _{cross}) \lesssim -0.5$ (see Fig.~\ref{fig:fbound}).  If
the bound fraction were larger over that range  of gas removal
time-scale to crossing time ratio, the shape of the ICMF would be
closer to that of the core mass function, while if $F_{\rm bound}$
were zero before increasing at, say, $m_{core} \simeq 10^5 - 3 \times
10^5\,{\rm M}_{\odot}$, this would result in a bell-shaped ICMF (see
the case of $\bar\varepsilon=0.20$ below).

The core mass function is most affected when $\bar\varepsilon=0.20$.
Fig.~\ref{fig:fbound} shows that the formation of a bound gas-free
star  cluster requires its parent core to be more massive than
$10^6\,{\rm M}_{\odot}$,  since the bound fraction is zero otherwise
(note that the top $x$-axis is  rightward-shifted by $\simeq 0.1$ in
$\log(\tau _{GR} / \tau _{cross})$ when $\varepsilon=0.20$, see
eq.~\ref{eq:ratio_tau}).  Owing to the wide variations of $F_{bound}$
over the core mass range ($0 \lesssim F_{\rm bound} \lesssim 0.85$),
the transformation of the core mass function into the ICMF is, in this
case, heavily core mass dependent: the featureless power-law core
mass function evolves into a bell-shaped ICMF.  The time-scale for this
evolution is that required for the exposed cluster to get back into virial 
equilibrium following gas expulsion, that is, 50\,Myr.  Examination of 
the model grid computed by \citet{bau07} for the SFE and the gas removal 
time-scale of relevance 
($\varepsilon =0.20$ and $\tau _{GR} \gtrsim \tau _{cross}$) shows that the
end of the violent relaxation phase occurs at about that age and the
cluster mass is then as defined in eq.~\ref{eq:minit}.

As indicated by the arrows in Fig.~\ref{fig:fbound}
and in the top panel of Fig.~\ref{fig:icmf}, the turnover location
is determined by the core mass at which the bound fraction is a
few per cent.  For instance, a core mass of $2\times10^6\,{\rm
M}_{\odot}$ with $\varepsilon =0.20$ leads to $\log (\tau
_{GR}/\tau _{cross})=0.13$ and  $F_{\rm bound}=0.06$.  The
corresponding initial cluster mass is thus $m_{\rm init}=(0.06) \times
(0.20) \times (2 \times 10^6\,{\rm M}_{\odot})  \simeq 2.4 \times
10^4\,{\rm M}_{\odot}$, roughly the mass of the ICMF turnover.

That $\varepsilon =0.20$ leads to a bell-shaped ICMF directly results
from the zero bound  fraction of stars for core masses less than
$\simeq 10^6\,{\rm M}_{\odot}$.  A SFE of $\varepsilon =0.25$ would 
result in the same effect (see Fig.~\ref{fig:fbound}).
This is equivalent to creating a
truncated power-law for the {\it cluster} forming cores,  even though
the {\it star} forming core mass function is a featureless power-law
down to  the low-mass regime.   That result is reminiscent of the
detailed investigation performed by \citet{par07} who found that a
system of cluster parent clouds devoid of low-mass objects, such as a
power-law mass function truncated at low-mass, results in a
bell-shaped ICMF.  We emphasize that, if these clusters were 
observed at an age of 50\,Myr, their observed mass would closely match
the mass predicted by our model.  Actually, all clusters of the bell-shaped
ICMF, irrespective of their mass, stem from gaseous 
progenitors more massive than $10^6\,{\rm M}_{\odot}$.  This implies
that unbound stars formed in the core leave the exposed cluster 
with velocities of order 60\,km.s$^{-1}$ and do not linger
as part of a halo around the bound part of the cluster.  
As a related consequence, note that if the very early Milky Way 
disc went through a violent star formation phase
producing such massive clusters, then the corresponding high velocity
dispersions lead naturally to the creation of a thick disc
component \citep{kroupa02}.

In the middle and bottom panels of Fig.~\ref{fig:icmf} we show  the
ICMF that results from the core mass function for SFE distributions
with variance $\sigma _{\varepsilon}=0.03$ and $0.05$ respectively.
The general features of the panels are the same as for zero variance,
however there are increasing differences, especially at the low-mass
end of the ICMF.  These differences are due to the effect of the SFE
not being symmetric around the mean $\bar\varepsilon$.  
For $\bar\varepsilon=0.2$, low-mass gas-embedded clusters with 
$\varepsilon=0.2$ are completely destroyed ($F_{\rm bound} \sim 0$) and
this is the case for any cluster selected below the mean.  However, a
low-mass gas-embedded cluster can be drawn with 
$\varepsilon = 0.3$~--~$0.4$, in which
case it will be able to survive and retain a significant fraction of
its mass.  For this reason, the ICMF flattens rather than
turning-over in the bottom panel of Fig.~\ref{fig:icmf}.

\begin{figure}
\begin{center}
\epsscale{1.2} \plotone{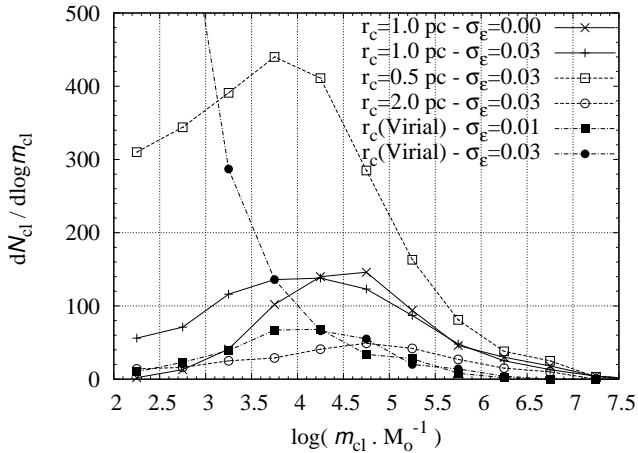}
\caption{How the shape and position of the ICMF produced by a $-2$
  power-law core mass function with a SFE of $\bar\varepsilon=0.20$,
  responds to variations in the width of the SFE distribution, and in
  the core radius}
\label{fig:to}
\end{center} 
\end{figure}

A key assumption underlying the above analysis is the hypothesis of a
constant radius for all star forming cores, regardless of their mass.
In Fig.~\ref{fig:to}, we examine the effect of the core radius
on the form of the ICMF produced from a core mass function with
$\varepsilon=0.20$ (ie. a bell-shaped ICMF).  We explore  how the
cluster mass at the turnover responds to core radius variations.  A
linear scaling for the $y$-axis is adopted as it enables us to
pinpoint the turnover location more accurately.  For the sake of
comparison, the ICMFs  obtained with $\sigma _{\varepsilon}=0.00$ and
$0.03$ and with $r_c=1$\,pc (dotted curves with crosses in top and
middle panels of Fig.~\ref{fig:icmf},  respectively) are shown as
solid lines.  ICMFs depicted with open symbols  correspond to constant
core radius of $r_c=0.5$\,pc and $r_c=2$\,pc.   A larger core radius
implies smaller core densities, thereby increasing the disruptive
effect of gas expulsion and lowering the number of bound gas-free star
clusters.  That the turnover is at higher cluster mass arises from the
leftward-shift of the mass-scaling of the  top $x$-axis of
Fig.~\ref{fig:fbound} when the core radius is larger  (see
eq.~\ref{eq:ratio_tau}).

As stated in Section \ref{sec:grts}, observations support the
assumption of an almost  constant core radius.  This, however, is at
variance with theoretical  expectations that the mass-radius relation
of virialised cores obeys  $r_c \propto m_c^{1/2}$
~\citep[e.g.][]{hp94}.  ICMFs obtained with  $r_c {\rm [pc]} = 10^{-3}
(m_c/1\,{\rm M}_{\odot})^{1/2}$  are shown as lines with filled
symbols in Fig.~\ref{fig:to}.  The normalisation is such that a core
of mass $10^6\,{\rm M}_{\odot}$ has a radius of 1\,pc.  Higher
normalisations, corresponding to less dense cores, lead to the
disruption of practically the whole original population of embedded
clusters.  While, for a constant core radius of  $r_c=1$\,pc, the
bell-shape of the ICMF is preserved when $\sigma _{\varepsilon}$ is
increased from 0.00 to 0.03, the virial mass-radius relation leads to
a power-law ICMF for $\sigma _{\varepsilon}=0.03$.  That is, the
sensitivity of the  ICMF shape to the width of the distribution
function for the SFE  is greater than in case of a constant core
radius.

\section{Summary}

The shape of the mass function of young star clusters remains much
debated.  Observational results are sorted into two distinct, often
opposed, categories: power-law or  bell-shaped mass functions.  We
have shown in this contribution that both shapes can arise from the
same physical process, namely, the supernova-driven gas expulsion from 
newly formed star clusters on a core mass dependent time-scale.
Specifically, more massive cores have a higher gas removal time-scale
to dynamical crossing-time ratio by virtue  of their deeper potential
wells (assuming the near constancy of core radii over the core mass
range).  For a given SFE, this results in larger bound  fractions
$F_{\rm bound}$ of stars at the end of the violent relaxation 
which follows gas expulsion.   The
shape of the ICMF is chiefly governed by the probability distribution
of the SFE  $\varepsilon$.  Bell-shaped ICMFs arise when most star
forming cores have  $\varepsilon \lesssim 0.25$.  In that case,
however, bound gas-free star clusters only  form out of cores more
massive than $10^6\,{\rm M}_{\odot}$.  This implies  that bell-shaped
ICMFs are also associated with gas-rich environments so as to
guarantee that the core  mass function is sampled to such a high mass.
In contrast, power-law ICMFs mirroring  the core mass function arise
when $\varepsilon \sim 0.4$ for most star forming cores.  This implies
that the shape of the ICMF also depends on the width  $\sigma
_{\varepsilon}$ of the SFE distribution.  When $\bar\varepsilon
\lesssim 0.25$, an increasing $\sigma _{\varepsilon}$ enhances the
contribution of cores with  $\varepsilon \simeq 0.40$ and turns a
bell-shaped ICMF into a power-law.   The amplitude of that evolution
is greater when core radii are assumed to follow  the mass-radius
relation of virialised cores than when the core radius is assumed  to
be constant.  As illustrated by Figs.~\ref{fig:icmf} and \ref{fig:to},
{\it that the mass function of young clusters observed in a variety of
environments appears not to be universal is not surprising}.

\citet{ves98} pointed out that a power-law ICMF of spectral index 
(-2) evolves to a bell-shaped mass function the turnover of which is located
at a significantly lower cluster mass than what is observed for the 
universal globular cluster mass function.  For such models, 
\citet{par05} showed that a power-law ICMF with $\alpha =-2$ also leads 
to a radial gradient in the peak of the evolved mass function
contrary to what is observed.  \citet{bau08} demonstrated that these two 
issues do not arise if the ICMF is depleted in low-mass clusters 
compared to a power-law of spectral index (-2) as a result of gas expulsion.  
In that case, it evolves within 13\,Gyr into a cluster mass function similar 
to that of old globular clusters both in the inner and outer Galactic halo 
(see their fig.~4).  While a mean SFE of $\bar\varepsilon \gtrsim 0.40$ preserves the shape of 
the core mass function, a mean SFE not exceeding $\bar\varepsilon \simeq 0.25$ 
leads to a low-mass cluster depleted ICMF (see Fig.~\ref{fig:icmf}).  
The universality of the globular cluster mass function 
may thus stem from the existence of an upper limit of, say, 25 per cent
on the mean SFE in all protogalaxies.  The origin of that constraint on the
mean SFE in the protogalactic era remains to be explained, however.

In this preliminary study, the energy input rate from massive stars encompasses 
the contribution made by supernovae only, i.e. that of stellar winds has been 
neglected.  Consequently, the above results apply to the sole case of metal-poor 
environments.  As noted in Section~2, we have assumed that the effect of gas 
expulsion depends on the actual star formation efficiency - i.e. we assume that 
the stars and gas (potential) are in virial equilibrium at the onset of gas 
expulsion.  This assumption is reasonable if (the stellar component of) the 
cluster has had time to relax.  A cluster will have been able to relax if the 
onset of gas expulsion occurs after a few crossing times.  In a low-metallicity 
regime such as we are considering in this paper, gas expulsion will be driven by 
supernovae rather than stellar winds and so will not begin until a few Myr after 
the stars have formed.  Such a time-scale is several crossing times, and so we 
could expect the eSFE to closely match the SFE.  We note however, that there 
may well be a mass-eSFE dependence in higher metallicity cluster populations 
that may alter our conclusions (which would be applicable to populations 
such as those in the Antennae merger of disc galaxies).

In a follow-up paper, we will consider the additional impact of 
stellar winds and the case of star clusters forming out of dense 
environments where strong tidal fields prevail.  



\acknowledgments
GP acknowledges support from the Alexander von Humboldt
Foundation in the form of a Research Fellowship and from
the Belgian Science Policy Office in the form of a Return Grant. 
GP and SG are grateful for research support and hospitality at the 
International Space Science Institute in Bern (Switzerland), 
as part of an International Team Programme. We acknowledge 
partial financial support from the UK's Royal Society through an 
International Joint Project grant aimed at facilitating networking 
activities between the universities of Sheffield and Bonn.







\end{document}